\documentclass[]{jfm}

\usepackage{bm} 
\usepackage{amssymb}
\usepackage{natbib}
\usepackage{graphicx}
\usepackage{color}
\usepackage{citesort}

\usepackage{amsmath}

\newcommand{\be}{\begin{equation}}
\newcommand{\ee}{\end{equation}}
\renewcommand{\r}{{\rho}}
\newcommand{\pt}{{\partial_t}}

\newcommand{\te}{\theta}

\newcommand{\half}{\frac{1}{2}}

\newcommand{\dt}{{\Delta t}}

\newcommand{\bea}{\begin{eqnarray}}
\newcommand{\eea}{\end{eqnarray}}

\usepackage{amsmath}
\usepackage{amsfonts}
\usepackage{amssymb}
\usepackage{bm}
\usepackage{color}
\usepackage{graphicx}

\begin{document}

\title[Lattice Boltzmann method with self-consistent  equilibria]{Lattice Boltzmann method with self-consistent thermo-hydrodynamic equilibria}

\author[M. Sbragaglia, R. Benzi, L. Biferale, H. Chen, X. Shan and S. Succi]
{M.\ns S\ls B\ls R\ls A\ls G\ls A\ls G\ls L\ls I\ls A$^1$, R.\ns B\ls E\ls N\ls Z\ls I$^1$,  L.\ns B\ls I\ls F\ls E\ls R\ls A\ls L\ls E$^{1}$, H.\ns C\ls H\ls E\ls N$^{2}$,  X.\ns S\ls H\ls A\ls N$^{2}$ and  S.\ns S\ls U\ls C\ls C\ls I$^{3}$}

\affiliation{$^1$Department of Physics and INFN, University of Rome Tor Vergata, \\ Via della Ricerca Scientifica 1, 00133 Rome, Italy \\
$^2$ EXA Corporation, 3 Burlington Woods Drive, Burlington, Massachussets 01803, USA \\$^3$ Istituto per le Applicazioni del Calcolo CNR, Viale del Policlinico 137, 00161 Roma, Italy}

\date{\today}
\maketitle

\begin{abstract}
  Lattice kinetic equations incorporating the effects of
  external/internal force fields via a shift of the local fields in
  the local equilibria, are placed within the framework of continuum
  kinetic theory.  The mathematical treatment reveals that, in order
  to be consistent with the correct thermo-hydrodynamical description,
   temperature must also be shifted, besides momentum.  New perspectives
  for the formulation of thermo-hydrodynamic lattice kinetic models of
  non-ideal fluids are then envisaged.  It is also shown that on the
  lattice, the definition of the macroscopic temperature requires the
  inclusion of new terms directly related to discrete effects. The
  theoretical treatment is tested against a controlled case with a non
  ideal equation of state.
\end{abstract}

\section{Introduction}

Lattice implementations of discrete-velocity kinetic models have
gained considerable interest in the last decade, as efficient tools
for the theoretical and computational investigation of the physics of
complex flows (\cite{SC1,SC3,Yeo,Waga,Wagb}).  An important class of
discrete-velocity models for ideal fluid flows, the lattice Boltzmann
models, (\cite{Gladrow,BSV,Chen}) can be derived from the continuum
Boltzmann (BGK) equation (\cite{BGK54}), upon expansion in Hermite
velocity space of the single particle distribution function, $f({\bm
  x}, {\bm \xi},t)$, describing the probability to find a molecule at
space-time location $({\bm x},t)$ and with velocity ${\bm \xi}$
(\cite{ShanHe98,Martys98,Shan06,Grad}).  As a result, the
corresponding lattice dynamics acquires a more systematic
justification in terms
of an underlying continuum kinetic theory.\\
The main goal of this paper is to extend such systematic link between
continuum and lattice formulation to the case of {\it hydrodynamical
  thermal fluctuations} in non-ideal fluids under the action of both
external (say, gravity) and/or internal forces.   In
particular, we shall show that if the effects of the force field are
taken into account via a uniform shift of the momentum in the equilibrium 
distribution, as proposed in \cite{SC1}, the evolution for the total kinetic energy  needs to be corrected as well.  A viable possibility is to introduce an {\it ad-hoc} shift in the temperature field entering  the local equilibrium. By doing so, the total kinetic energy recovers the correct  thermo-hydrodynamical evolution.\\
Let us consider the usual continuum, $D$-dimensional, Boltzmann
BGK equation \be\label{LBM} \frac{\partial f}{\partial t}+{\bm \xi}
\cdot {\bm \nabla} f+{\bm g}\cdot {\bm \nabla}_{\xi}
f=-\frac{1}{\tau}\left(f-f^{(0)} \right);\qquad \; f^{(0)}({\bm \xi};
\rho, \theta, {\bm u})=\frac{\rho}{(2 \pi \theta)^{D/2}} e^{-|{\bm
    \xi}- {\bm u}|^2/2 \theta} \ee
where ${\bm g}$ represents an acceleration field and $\tau$ a  relaxation time towards the local equilibrium $f^{(0)}$. This local equilibrium will depend on the local density, $\rho$, momentum, $\rho {\bm u}$, and temperature, $\theta$.\\
We will show that, as far as the macroscopic evolution of
  the hydrodynamical fields is concerned, it is possible to
renormalize the action of the force $\rho {\bm g}$ only in terms of a
suitable {\it shift} of the local Maxwellian equilibrium distribution
appearing in (\ref{LBM}): $f^{(0)}({\bm \xi}; \rho, \theta, {\bm u})
\rightarrow {\bar f}^{(0)}({\bm \xi}; \rho, {\bar \theta}, {\bm {\bar
    u}})$.  The new --shifted-- Boltzmann formulation being:
\begin{eqnarray} \label{MASTERSHIFT3}
\frac{\partial f }{\partial t}+{\bm \xi} \cdot {\bm \nabla} f=-\frac{1}{\tau}(f- {\bar f}^{(0)}); \qquad   \; 
{\bar f}^{(0)}({\bm \xi}; \rho, \theta, {\bm u})=\frac{\rho}{(2 \pi {\bar \theta})^{D/2}} e^{-|{\bm \xi}- {\bm {\bar u}}|^2/2  {\bar  \theta}}.
\end{eqnarray}
The first result of the paper is to show that, in order to recover the
correct thermo-hydrodynamic behaviour, the shifted local velocity and
temperature must take the following form:
\begin{equation}
\label{eq:shift2}
\bar {\bm u} = {\bm u} + \tau {\bm g} \qquad \bar \theta = \theta -  \tau^2 g^2/D.
\end{equation}
The idea of shifting momentum has been pioneered for the case 
of a driving force due to internal, self-consistent interactions,
which depend on the system configuration, typically the density field
distribution.  Such kind of extension has proven instrumental to the
succesfull formulation of lattice kinetic theory of isothermal
non-ideal fluids (\cite{SC1,SC3}), in which one is concerned with the
 isothermal hydrodynamic evolution of density and momentum alone.

Here we extend it further to the important case
of thermal hydrodynamic fluctuations.  Indeed, the use of shifted-equilibria has
many immediate and important methodological consequences: (i) it
provides an elegant way to incorporate the force effects thereby
dispensing with the need of taking derivatives of the distribution
function in velocity space; (ii) it allows a systematic derivation of
lattice kinetic equations for non-ideal fluids with (pseudo)-potential
energy interactions (\cite{Shan06,Martys98,Guo02}); (iii) it
highlights the need of including a suitable redefinition of the
hydrodynamical fields on the lattice, in order to recover the correct
continuum limit of the thermo-hydrodynamical equations for density,
momentum and total kinetic energy, ${\cal K}$.  Here and throughout,
by continuum thermo-hydrodynamic limit we refer to the following set
of macroscopic equations (repeated indices are summed upon):
\begin{equation}\label{eq:1}
\begin{cases}
\partial_t \rho + \partial_i ( \rho u_i) = 0  \\ 
\partial_t (\rho u_k) + \partial_i (P_{ik}) = \rho g_k  \\ 
\partial_t {\cal K}    + \frac{1}{2} \partial_i q_i = \rho g_i u_i
\end{cases}
\end{equation}
where $P_{ik}$ and $q_i$ are momentum and energy fluxes (still unknown
at this level of description).  More precisely, we shall show that the
above equations can be obtained {\it exactly} from the previous
continuum Boltzmann Equation with shifted equilibrium
(\ref{MASTERSHIFT3}). 
This would not be the case, if the momentum is the only
shifted quantity in (\ref{MASTERSHIFT3}). 
The
relevance of such a continuum kinetic theory is mainly motivated by
the final goal of formulating lattice versions of the Boltzmann
Equations for non-ideal fluids, including thermo-hydrodynamic
effects.  In fact, in lattice formulations, the need of representing
velocity degrees of freedom through a limited set of discrete speeds,
raises the problem of a correct and efficient implementation of the
continuum velocity-streaming operator ${\bm g} \cdot {\bm \nabla}_\xi
f$.  We shall show that shifted equilibria in the continuum
representation have a well-defined lattice analogue, so that the
lattice counterpart of the continuum description (\ref{MASTERSHIFT3})
can be obtained through the usual lattice Boltzmann discretization:
\be f_{l}({\bm x}+ {\bm c}_{l} \dt,t+\dt) - f_{l}({\bm x},t) =
-\frac{\dt}{\tau}(f_{l}({\bm x},t) -{f}_{l}^{(0)}({\bm
  x},\rho^{(L)},\bar{\bm
  u}^{(L)},\bar\theta^{(L)})) \label{eq:lbe.lat} 
\ee 
where the subscript $l$ runs over the discrete set of velocity on the lattice,
${\bm c}_l$, and the superscript $L$ indicates that the macroscopic
fields are now defined in terms of the lattice Boltzmann populations:
$$ \rho^{(L)}  = \sum_l f_l; \qquad \rho^{(L)}  {\bm u}^{(L)}   = \sum_l {\bm c}_l f_l; \qquad D \rho^{(L)} \theta^{(L)}   = \sum_l |{\bm c}_l - {\bm u}^{(L)}|^2 f_l. $$
In the expression (\ref{eq:lbe.lat}), the Boltzmann equilibrium (see \cite{Nie08} for its explicit expression) is 
computed with shifted momentum and temperature, as follows:
$$\bar{\bm u}^{(L)}= {\bm u}^{(L)} + \tau {\bm g} \qquad \bar{\theta}^{(L)}= \theta^{(L)} +\Delta \theta. $$
After some algebra, it can be shown that the temperature shift $\Delta \theta$ 
can be expressed in closed form as a function of the lattice time-step $\Delta t$:
$$
\Delta \theta = \frac{\tau(\dt-\tau) g^2}{D } + {\cal O} (\Delta t)^2 + \dots
$$
Moreover, in order to recover the thermo-hydrodynamical equations 
\begin{equation}
\begin{cases}
\pt \r^{(H)} + \partial_i ( \r u_i^{(H)}) = 0\\
\pt (\r^{(H)} u_k^{(H)}) + \partial_i (P_{ik}^{(H)}) = \rho^{(H)} g_k\\
\pt  {\cal K}^{(H)} + \half \partial_i q_i^{(H)} = \rho^{(H)} g_i u_i^{(H)}.
\end{cases}
\label{eq:hydro}
\end{equation}
the hydrodynamical fields can be computed in terms of a closed expansion to all orders in $\Delta t$ and  can be calculated in terms of a suitable lattice operator.  For example, density is left unchanged, $\rho^{(H)} = \rho$, while the first non trivial correction  is given by the well-known pre and post-collisional momentum average (\cite{Buick00}):
$$
{\bm u}^{(H)}={\bm u}^{(L)}+\frac{\Delta t}{2} {\bm g}
$$
as well as by a new, non-trivial, correction to the temperature field:
$$
{\theta}^{(H)}  = \te^{(L)} + \frac{(\dt)^2g^2}{4 D }.
$$
First, we notice that in the limit $\dt \rightarrow 0$, the lattice formulation for both
shifted fields and hydrodynamical fields goes back to the continuum one, as it should.
Second, the continuum formulation in terms of shifted fields
indicates a straightforward link with the discrete variables via the
Hermite-Gauss expansion (\cite{Shan06}). 
Third, and maybe more important for applications, we emphasize that in order 
to achieve a self-consistent thermo-hydrodynamical description in the lattice (\ref{eq:hydro}), 
both momentum --as it was known-- and temperature, acquire discrete corrections. \\
 
The paper is organized as follows: in section \ref{SEC1} we present the basic ingredients of the continuum BGK model  with shifted equilibria and we show how to compute the correct normalization of the local fields, so as to recover 
the correct macroscopic equations. 
This procedure is extended to the lattice models in section \ref{SEC2} with a detailed analysis of the discrete contribution to the renormalization procedure. A numerical test of the above arguments is provided in section \ref{SEC4}. Finally, in section \ref{SECH}, we will briefly discuss the physical interpretation of the shifted continuum model and propose some further development.


\section{Shifted Continuum Equilibrium}\label{SEC1}

In this section we deal with the macroscopic properties of a continuum
model (\ref{MASTERSHIFT3}). 
The main goal is  to renormalize the effects of the forcing term ${\bm g}\cdot {\bm \nabla}_{\xi}f$  in (\ref{LBM}) via a suitable local  equilibrium
with shifted fields:
\begin{equation}\label{SHSH}
\bar {\bm u} = {\bm u} + \Delta {\bm u}({\bm g}, \tau); \qquad \bar \theta = \theta +\Delta{\theta}({\bm g}, \tau).
\end{equation}
It is well known (\cite{BGK54,Gladrow}) that the usual definition of Boltzmann Equation with explicit forcing  given in
(\ref{LBM}) leads to the exact macroscopic equations (\ref{eq:1})
with the averaged  fields given by \be
\rho = \int d \xi f; \qquad  \rho {\bm u} = \int d \xi {\bm \xi} f ;  \qquad 
{\cal K}=\frac{1}{2}(\rho D \theta + \rho u^2) = \frac{1}{2} \int d \xi \xi^2 f.  \label{macro}
\ee
The momentum and energy fluxes
$$
P_{ij}=\int d \xi \xi_i \xi_j f \hspace{.2in} q_{i}=\int d \xi  \frac{\xi^2}{2} \xi_i  f 
$$
in the LHS of equations (\ref{eq:1}) need to be closed, a task which
is usually accomplished via the Chapman-Enskog expansion. In order to
derive (\ref{eq:1}) from (\ref{LBM}) it is indeed sufficient to notice
that the collision operator $-\frac{1}{\tau}(f-f^{(0)})$ preserves
both mass, momentum and total kinetic energy, as long as the local
equilibrium is expressed in term of the macroscopic fields, $\rho,{\bm
  u}, \theta$, i.e.  whenever, besides the definition (\ref{macro}),
the local equilibrium carries the same mass, momentum and energy of
the actual distribution, namely: \be \rho = \int d \xi f^{(0)}; \qquad
\rho {\bm u} = \int d \xi {\bm \xi} f^{(0)} ; \qquad {\cal
  K}=\frac{1}{2}(\rho D \theta + \rho u^2) = \frac{1}{2} \int d \xi
\xi^2 f^{(0)}.  \label{macro0} 
\ee 
When considering the effect of the
forcing field embedded within the shifted equilibrium
(\ref{MASTERSHIFT3}), the only difference with the previous standard
derivation comes from the fact that now the averaged macroscopic
fields, when evaluated on the {\it shifted} equilibrium, {\it do not }
coincide with the hydrodynamical fields defined in terms of the local
particle distribution (\ref{macro}), i.e. the collision operator,
$-\frac{1}{\tau}(f-\bar{f}^{(0)})$ preserves momentum and energy only
globally, but not locally. Still, it is easy to realize that the extra
momentum and energy brought by the shifted equilibrium is given by:
\be \label{macro1} \rho \bar {\bm u} - \rho {\bm u} = \int d \xi {\bm
  \xi} ({\bar f}^{(0)} - f) ; \qquad \frac{1}{2}(\rho D \bar \theta +
\rho \bar u^2) - \frac{1}{2}(\rho D \theta + \rho u^2) = \frac{1}{2}
\int d \xi \xi^2( {\bar f}^{(0)} - f) \ee and that if we chose the
shifted fields as given by expressions (\ref{eq:shift2}) the exact
macroscopic equations (\ref{eq:1}), for density, momentum and kinetic
energy evolution, are recovered.  To this purpose, it is sufficient to
evaluate the first three lowest momenta of eq. (\ref{MASTERSHIFT3})
and use the relations (\ref{macro1}) and (\ref{macro}).
This is the first result of the present work. \\
Let us stress once again that the shift in the
temperature only responds to the {\it need} of cancelling out extra terms 
in the rhs of eq. (\ref{eq:1}), that would otherwise result from the momentum-shift alone. One may wonder if beside the formal correct unclosed equations
(\ref{eq:1}) the two Boltzmann Equations formulation (\ref{LBM}) and
(\ref{MASTERSHIFT3}) do also share the same hydrodynamical behaviour,
i.e. if the unclosed momentum and heat fluxes do have the same
closure.  By performing the whole Chapman-Enskog expansion, it can be
shown that this is indeed the case, at least up to second order in the
expansion parameter where dissipative terms in momentum and heat
appear. Details of these calculations will be reported elsewhere.
\section{Lattice Implementation}\label{SEC2}
In this section, we treat the lattice averaged equations and discuss the way that the
corresponding shifts in the momentum and temperature fields are affected by the discretization of the algorithm. 
To this purpose, let us go back to the lattice Boltzmann equation with shifted momentum and temperature given
by expression (\ref{eq:lbe.lat}).  
First, we Taylor expand the LHS of (\ref{eq:lbe.lat}), and obtain the lattice-series expression
$$
D_{l,t} f_l+\frac{\dt}{2} D_{l,t}^2 f_l+\frac{(\dt)^2}{6} D_{l,t}^3f_l+...= -\frac{1}{\tau}\left( f_l -f_l^{(0)} \right). 
$$
with $D_{l,t}=\pt+{\bm c_l} \cdot {\bm \nabla}$. 
We can then rewrite (\ref{eq:lbe.lat}) in a compact form
\be \label{eq:comp}
\left( e^{\dt D_{l,t}} -1 \right) f_l=\dt  {\cal C}_l
\ee
with ${\cal C}_l=-\frac{1}{\tau} \left(f_l-f^{(0)}_l \right)$ the collisional operator. 
A formal inversion of (\ref{eq:comp}) yields:
$$
D_{l,t} f_l= \frac{\dt D_{l,t}}{\left( e^{\dt D_{l,t}} -1 \right)} {\cal C}_l = {\cal C}_l+\left[\frac{\dt D_{l,t}e^{-\dt D_{l,t}}}{\left( 1- e^{-\dt D_{l,t}} \right)} -1  \right]{\cal C}_l
$$
where the LHS is recognized as the generating function of Bernoulli polynomials $B_n(x)$ (\cite{GradR}) : 
$$\frac{e^{xt}}{e^t-1}=\sum_{n=0}^{\infty} B_n(x) \frac{t^{n-1}}{n!}.$$
Let us also introduce the operator $ {\cal L}_{l,\dt}$, defined by the Taylor expansion in $D_{l,t} \Delta t$:
$$
D_{l,t} {\cal L}_{l,\dt}=\left[\frac{\dt D_{l,t}e^{-\dt D_{l,t}}}{\left( 1- e^{-\dt D_{l,t}} \right)} -1  \right]=-\frac{\dt D_{l,t}}{2}\left(1- \frac{\dt D_{l,t}}{6}+\frac{(\dt D_{l,t})^3}{360}-\frac{(\dt D_{l,t})^5}{15120} +{\cal O}\left( (\dt D_{l,t})^7 \right)\right).
$$
The relevant point is that the above operator can be rewritten in terms of {\it a  lattice operator} 
performing an inverse shift in space and time:
$S_{l,\Delta t}=(e^{-\dt D_{l,t}}-1)$. 
It is easy to realize that its action on any field, say $\phi$, defined on the lattice gives back:
$$
S_{l,\Delta t} \phi({\bm x},t) = \phi({\bm x} - \dt {\bm c}_l, t -\dt).
$$
The action of the operator ${\cal L}_{l,\dt}$ can be recast as follows:
$$
{\cal L}_{l,\dt}=-\frac{\dt}{2}\left(1+ \frac{1}{6} S_{l,\Delta t}- \frac{1}{12}S_{l,\Delta t}^2+ \frac{19}{360}S_{l,\Delta t}^3 -\frac{3}{80}S_{l,\Delta t}^4+{\cal O}\left(S_{l,\Delta t}^5 \right) \right).
$$
This shows that it is possible to rewrite the final dynamics, exactly to all orders in $\dt$, by 
retaining {\it only} shift-operators on the lattice:
$$
D_{l,t} \left(f_l -{\cal L}_{l,\dt} {\cal C}_l \right)={\cal C}_l
$$
Alternatively, by recalling the definition of ${\cal C}_l$, we also have:
$$
D_{l,t}  \left(f_l +\frac{1}{\tau}{\cal L}_{l,\dt}\left( f_l-{f}^{(0)}_l \right)  \right)=-\frac{1}{\tau}(f_l-f_l^{(0)}).
$$
We can then average this equation in velocity space and look at the equations for the first three-order momenta, so
as to recover the hydrodynamic evolution for density, momentum and total energy (\ref{eq:hydro}).
Simple calculations shows that the  LHS set of macroscopic equations (\ref{eq:hydro}) 
is obtained by  means of the following definitions of macroscopic hydrodynamic ($H$) fields 
(scalars, vectors and second-order tensors)
$$
\begin{cases}
\rho u_i^{(H)}=\sum_l c^i_l f_l+ \frac{1}{\tau}\left( \sum_l c^i_l {\cal L}_{l,\dt}[f_l-{f}_l^{(0)}] \right)\\
P_{ij}^{(H)}=\sum_l c^i_l c^j_l f_l+\frac{1}{\tau}\left( \sum_l c^i_lc^j_l {\cal L}_{l,\dt}[f_l-{f}_l^{(0)}] \right)\\
{\cal K}^{(H)}=\left( \frac{D}{2} \rho \theta^{(H)}+\frac{1}{2} \rho (u^{(H)})^2 \right)=\frac{1}{2}\sum_l c_l^2 f_l+ \frac{1}{\tau}\left( \frac{1}{2}\sum_l c_l^2 {\cal L}_{l,\dt}[f_l-{f}_l^{(0)}] \right)\\
q_{i}^{(H)}=\frac{1}{2}\sum_l c_l^2c^i_l f_l+ \frac{1}{\tau}\left( \frac{1}{2}\sum_l c_l^2c^i_l {\cal L}_{l,\dt}[f_l-{f}_l^{(0)}] \right).
\end{cases}
$$
In order to capture the correct RHS of (\ref{eq:hydro}) as well, we must choose the fields entering the shifted equilibrium, $f^{(0)}$ 
in (\ref{eq:lbe.lat}),  $\bar{\bm u}^{(L)}={\bm u}^{(L)}+\Delta {\bm u}$, $\bar{\theta}^{(L)}=\theta^{(L)}+\Delta \theta$,
in the following form:
$$
\begin{cases}
\Delta {\bm u}=\tau  {\bm g}\\
\Delta \theta=\frac{2 \tau}{\rho D} \left(\rho g_i (u^{(H)}_i-u_i)-\frac{\tau \rho g^2}{2} \right).
\end{cases}
$$
Let us notice that the above expression for the temperature shift is implicit, i.e. it is given in terms of the hydrodynamical velocity $u^{(H)}$ which depends itself on the equilibrium.  One may get a closed expression only via a Taylor expansion in $\dt$.  For example, to first order in the expansion of ${\cal L}_{l,\dt}$, we have simply 
$$
{\cal L}_{l,\dt}=-\frac{\Delta t}{2}
$$
corresponding to the following temperature shift: 
$$\bar{\theta}^{(L)}  = \te^{(L)} + \frac{\tau(\dt-\tau) g^2}{D }
$$
Consequently, the hydrodynamical velocity and temperature become:
$$
\begin{cases}
{\bm u}^{(H)}={\bm u}^{(L)}+\frac{\Delta t}{2} {\bm g}\\
\theta^{(H)}  = \te^{(L)}  + \frac{(\dt)^2 g^2}{4  D}.
\end{cases}
$$
where the hydrodynamic velocity is nothing but the pre and
post-collisional average, while the non trivial correction in $\dt$ to
the temperature fields is the new result, as anticipated in the
introduction.

\section{Numerical test}\label{SEC4}
We now proceed to the implementation of a test case of the above procedure, where the
need for the temperature shift appears in full.  
The most important instance where total energy conservation is crucial is
the case of a gas(fluid) departing from ideal conditions, as a result of
an internal, self-consistent, potential. In this case, the
thermo-hydrodynamical equations must conserve the total energy, given
by the sum of the total kinetic plus the potential energy. 
Typical relaxation experiments will then show a non trivial
exchange between the kinetic and potential energy components, until
a dynamical or static equilibrium is finally attained.

We specialize the discussion to a simple, and yet non-trivial, case where
the interparticle force is purely {\it repulsive}
:\be\label{FORC} {\bm F}=\rho {\bm g}({\bm x},t)={\cal G}
\rho({\bm x},t) \sum_l w_l \rho({\bm x}+{\bm c}_l\Delta t,t){\bm c}_l
\approx -{\bm \nabla} P_{b}^{(int)} \hspace{.2in}
P_{b}^{(int)}(\rho)=-\frac{1}{2} {\cal G} \rho^2 \ee with $w_l$ a
suitable set of weights which enforce the right symmetries on the
lattice (\cite{SC1,SC3}). Repulsion is imposed by choosing a negative coupling
constant, ${\cal G} \le 0$.  This case allows full control of
the non-ideal part of the equation of state.  Indeed, the bulk
pressure provides the usual ideal-gas contribution, $P^{(id)}_b=\rho \theta$
plus the non-ideal one, given by the Taylor expansion of the forcing
term in (\ref{FORC}),  $P^{(tot)}_b = P^{(id)}_b+P^{(int)}_b$. 
The system also has internal potential energy $E_{V}=-\frac{1}{2} {\cal G} \rho^2$, where we have neglected possible contributions coming from strong density gradients. The transport equation for this intermolecular potential energy (\cite{HeDoolen01,Snidera,Sniderb}) reads as follows:
$$
\partial_t E_{V}+\partial_{k}(u^{(H)}_k E_{V})=-(\partial_j u^{(H)}_j)
P_{b}^{(int)},
$$
as can be readily derived from the density evolution in (\ref{eq:1}). By summing to the total kinetic energy evolution, we obtain the following total energy balance:
$$
\partial_t (E_V+{\cal K}^{(H)})=-\partial_j( u^{(H)}_j P_{b}^{(int)})-\frac{1}{2}\partial_iq_i^{(H)}-\partial_{k}(u^{(H)}_k E_V),
$$
whose divergence-like structure at the RHS, ensures total energy conservation.\\ 
Let us stress that the reconstruction of a total divergence  is only possible thanks to the superposition of the contribution $(\partial_j u^{(H)}_j) P_{b}^{(int)}$, stemming from 
the evolution of the potential energy, plus the contribution $u^{(H)}_j (\partial_j P_{b}^{(int)})$, coming from the RHS 
of the total kinetic energy in (\ref{eq:hydro}). Here, we fully appreciate the importance of 
the temperature shift, in order to recover the correct total energy dynamics. Using fully periodic boundary
conditions, the shifted lattice
Boltzmann formulation is therefore expected to provide conservation 
of the total energy from the hydrodynamical point of view.

\begin{figure}
\begin{center}
\includegraphics[scale=0.7]{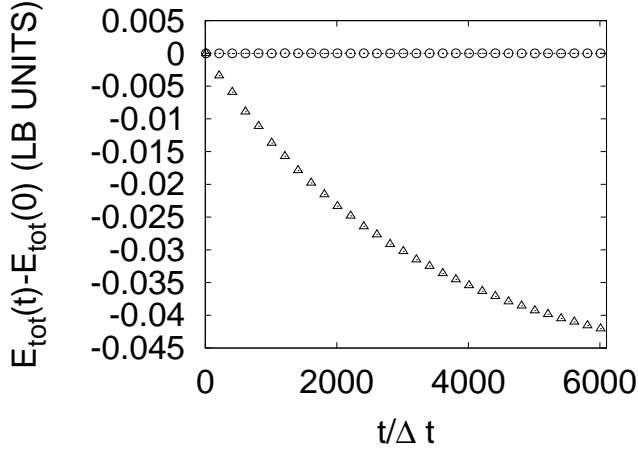}
\caption{Variations of total energy for a non ideal system with an
  initial sinusoidal wave in the temperature field
  $\theta^{(H)}(x,y,t=0)=1.0+\epsilon \sin(2 \pi y /L_y)$ with
  $\epsilon=0.01$ and smooth hydrodynamical velocity fields. The
  lattice Boltzmann parameters in (\ref{FORC}) and (\ref{eq:lbe.lat}) are
  such that $\tau/\Delta t=0.6$, ${\cal G}=-3.0$. Two simulations are
  carried out. The first simulation is only with shifted momentum
  ($\triangle$). A second one with shifted momentum and temperature
  ($\circ$). }
\label{fig:1}
\end{center}
\end{figure}

To ensure a sufficiently accurate recovery of the thermal transport
phenomenon, we employ a two-dimensional $37$-speed LB model, corresponding to a ninth-order accurate Gauss-Hermite quadrature. In
conjunction, the following fourth-order Hermite expansion of the
Maxwellian is used as an equilibrium distribution
(\cite{Shan06,Nie08}).  A simulation is performed on a $L_x \times L_y$
= $10 \times 100$ grid, with a small perturbation of a single
sinusoidal wave in the temperature field
$\theta^{(H)}(x,y,t=0)=1.0+\epsilon \sin(2 \pi y /L_y)$
($\epsilon=0.01$). The initial density field is constant.  The
difference of the total energy $\int ((E_V+{\cal K}^{(H)}))(t) dx dy $
with respect to its initial value is monitored in figure
\ref{fig:1}. This figure clearly shows that the lattice Boltzmann formulation
without shifted temperature is not able to sustain satisfactory energy
conservation. On the other hand, upon shifting the temperature
field, the correct energy balance is recovered. The energy conservation
is still below machine precision, due to the fact that our expression for the total energy 
is given in terms of a continuum description of the non-ideal forcing term (\ref{FORC}). In order to further improve the accuracy of energy conservation, a discrete version of the internal potential energy, $E_V$, thermodynamically consistent on the lattice, needs to be developed. 


\section{Conclusions and Perspectives}\label{SECH} 

Before concluding, let us discuss further the physical meaning of the
shifted Boltzmann Equation (\ref{MASTERSHIFT3}). It is easy to realize
that it can also be rewritten as:
\begin{eqnarray}\nonumber
\frac{\partial f ({\bm x},{\bm \xi},t)}{\partial t}+{\bm \xi} \cdot {\bm \nabla} f({\bm x},{\bm \xi},t)=-\frac{1}{\tau}\left(f({\bm x},{\bm \xi},t)-e^{-\tau {\bm g} \cdot {\bm \nabla_{\xi}} -\tau^2 \frac{g^2}{2D} {\bm \nabla_{\xi}}\cdot {\bm \nabla_{\xi}}   } f^{(0)}({\bm \xi}; \rho, \theta, {\bm u}) \right).
\end{eqnarray}
Upon Taylor expanding the RHS terms corresponding to the shift in the momentum, $-e^{-\tau {\bm g} \cdot {\bm \nabla_{\xi}}} f^{(0)}$, up to second order in $\tau$,
we obtain: 
\be\label{FILTERED}
\frac{\partial f}{\partial t}+{\bm \xi} \cdot {\bm \nabla} f=-\frac{1}{\tau}\left(f-f^{(0)} \right) -{\bm g}\cdot {\bm \nabla}_{\xi} f^{(0)} +\frac{\tau}{2} {\bm g}{\bm g} : {\bm \nabla}_{\xi} {\bm \nabla}_{\xi} f^{(0)}
\ee
It is simple to check that, as for the isothermal dynamics of density and momentum, the shifted equilibrium
Boltzmann equation (\ref{MASTERSHIFT3}) is equivalent to (\ref{FILTERED}). 
A stabilizing diffusion term in velocity space stands therefore out.\\
It is also noted that, as a first order of approximation, this kind of diffusion term can be thought as deriving
from a standard BGK dynamics, with the equilibrium distribution $f^{(0)}({\bm x},{\bm \xi},t)$ replaced by a 
smoothed version, resulting from coarse-graining in velocity space, filtering 
fluctuations up to  $\delta \bm v^{\prime} < \bm g \tau$. Considering  the Taylor expansion of the temperature shift operator,
$-e^{ -\tau^2 \frac{g^2}{2D} {\bm \nabla_{\xi}}\cdot {\bm \nabla_{\xi}}   } f^{(0)}$,
we obtain, up to $\tau^2$,  an extra-term proportional to: \be \label{eq:tem} g^2 \Delta_{\xi} f^{(0)}.\ee
Such contribution can be interpreted as deriving from a stochastic component in the
acceleration field. To illustrate the point, let us start again from the
continuum BGK equation (\ref{LBM}) and let us consider the streaming
term in velocity space, $\frac{d {\bm \xi}}{d t}\cdot {\bm
\nabla}_{\xi} f$, in which the molecular velocity $\xi_i$ obeys the following Langevin equation:
$$
\frac{d \xi_i}{d t}=\eta_i
$$
where $\eta_i$ is a standard delta-correlated Gaussian noise with zero average and the following normalization:
$$\langle \eta_i \rangle=0 \hspace{.2in} \langle \eta_i \eta_j \rangle=  \frac{1}{2} \sigma \delta_{ij}. $$ 
We expect the stochastic term to provide a mechanism for producing thermal fluctuations in the fluid.  
The key ingredient is a correct evaluation of the term
$
\langle \frac{d \xi_i}{dt} \nabla_{\xi_i} f  \rangle
$
where $\xi_i$ is now a stochastic variable and $\langle ... \rangle$ stands for an average over possible realizations of the stochastic term.  
We can apply Novikov's theorem (\cite{Novikov}) and approximate  $\hspace{.03in}\xi_i \approx \frac{d \xi_i}{d t} \tau=\eta_i \tau$. 
This yields:
$$
\langle \frac{d \xi_i}{dt} \nabla_{\xi_i} f  \rangle = \langle \frac{d \xi_i}{dt} \xi_j  \rangle \langle \nabla_{\xi_i} \nabla_{\xi_j} f \rangle \approx \tau \langle \eta_i \eta_j  \rangle \langle \nabla_{\xi_i} \nabla_{\xi_j} f \rangle
$$
which is of the same form of (\ref{eq:tem}) for a suitable choice of $\sigma$.
Let us emphasize that the shifted equilibrium formulation
(\ref{MASTERSHIFT3}) is {\it equivalent} to the standard BGK
formulation (\ref{LBM}), as far as the macroscopic equations
(\ref{eq:1}), and their Chapman-Enskog expansion up to second order,
are concerned. The diffusive extra-terms stemming from the Taylor
expansion indicate that, although maintaining the same hydrodynamics,
the shifted equilibrium formulation should nonetheless
feature better stability with concern to the global evolution of the
probability density, $f({\bm x}, {\bm \xi},t)$.


Summarizing, we have investigated lattice kinetic equations
incorporating the effects of external/self-consistent force fields via
a shift of the local fields in the local equilibria.  The mathematical
treatment reveals that, besides momentum, temperature also receives a
self-consistent shift from the force field. The  contribution of the temperature shift can also be traced back to a             {\it stochastic} component in the acceleration field, thus 
pointing to potentially new directions for the formulation of lattice Boltzmann
models for non-ideal fluids with thermo-hydrodynamic transport effects
(\cite{Brennen,Rowlinson,Degennes}).  Work along these lines is in
progress, which will hopefully permit to attack a broad class of
complex flow problems with thermal effects, such as thermally driven
phase-transitions, crystal growth, melting and many other
non-equilibrium thermo-hydrodynamic transport problems.

Finally, it has been shown that, in order to recover the correct
thermo-hydrodynamical equations (\cite{Bogho}) on the lattice, the
macroscopic temperature must acquire new terms, directly related to
the lattice spacing.  These terms naturally vanish in the continuum
limit, thus preserving the consistency of the discrete theory.  Many
items remain open to future investigation.  For instance, it would be
interesting to extend the present treatment to more general collision
operators, including multi-time relaxation models (\cite{Dellar}), which
would permit to model fluids at non-unitary Prandtl numbers
(\cite{Lohse}). The establishment of a H-theorem for
continuum and discrete kinetic equations with self-consistently
shifted equilibria, also appears an interesting topic for future
research (\cite{Karlin,Ansumali})

\end{document}